\documentclass[9pt,conference]{IEEEtran}
\usepackage{xcolor}
\usepackage{graphicx}
\usepackage{color}
\usepackage{placeins}
\usepackage{flushend}
\usepackage{svg}
\usepackage{float}
\usepackage{tabularx,colortbl}
\usepackage{graphicx}   
\usepackage{amsmath}   
\usepackage{mathtools,amssymb}
\usepackage[caption=false,font=footnotesize]{subfig}
\usepackage{siunitx}
\usepackage{bm}
\usepackage{cite}
\usepackage[skins,theorems]{tcolorbox}
\usepackage{empheq}
  {\empheq[box=\tcbhighmath]{align*}}
  {\endempheq}
  
\providecommand*{\mrm}[1]{\mathrm{#1}}

\renewcommand{\vec}[1]{{\boldsymbol#1}}

\providecommand*{\degree}{\ensuremath{^\circ}}

\newcommand{\ts}{\textsuperscript}
\newcommand{\clight}{\mrm{c}_0}

\newcommand{\eg}{\textit{e.g.}\/, }

\newcommand{\rev}[1]{\textcolor{black}{#1}}
\begin{document}
\title{Inverse Synthetic Aperture Radar, Radar Cross Section,  and Iterative Smooth Reweighting $\ell_1$-minimization}
\author{\IEEEauthorblockN{
        Christer Larsson\IEEEauthorrefmark{1}\IEEEauthorrefmark{2}
    }
  \IEEEauthorblockA{\IEEEauthorrefmark{1}
Saab Dynamics, SE-581 88 Linköping, Sweden}
\IEEEauthorblockA{\IEEEauthorrefmark{2}
Lund University, P.O. Box 118, SE-221 00 Lund, Sweden}
}
\maketitle
\begin{abstract}
    Radar Cross Section measurement data is often analyzed using Inverse Synthetic Aperture Radar images. This paper compares backprojection and iterative smooth reweighted $ \ell_1 $-minimization as methods to analyze radar cross section measurements and extract radar cross section for parts of the measured object. The main conclusion is that using backprojection images to extract RCS is robust and accurate but is more limited by the resolution than iterative smooth reweighted $ \ell_1 $-minimization. The latter method can be used for closely spaced scatterers but is limited in accuracy.
\end{abstract}

\section{Introduction}
Inverse Synthetic Aperture Radar (ISAR) can be used to make an image containing the location and complex valued intensity of the different scattering contributions from the measured target using the data from a Radar Cross Section (RCS) measurement. ISAR is therefore a very useful and often used tool when analyzing high resolution RCS measurements. The basics of RCS and ISAR are described in \eg \cite{Knott2004+etal,Mensa1991}. Downrange resolution is obtained by acquiring complex valued data over a frequency range and crossrange resolution is obtained by acquiring data over an angular range.  The object is placed on a rotating turntable in the ISAR measurement setup that we consider in this paper. The RCS is then measured and an inverse problem is solved to obtain an image.

It is common to use back projection as a straightforward method for ISAR but it comes with inherent limitations in resolution and ambiguity \cite{Vaupel+Eibert2006,Larsson2014,Larsson+Gallstrom2022,Larsson2024}.
The work presented in this paper uses an $ \ell_1 $-minimization method combined with  iterative smooth reweighting (ISR) \cite{Candes+etal2008,Pinchera+etal2019,Pinchera+Migliore2021,Larsson+Gallstrom2022,Larsson2024}.
We present and discuss the pros and cons of these methods  \eg in terms of resolution and accuracy. 
\section{Theory}
\label{sec:theory}
 The system of equations to be solved for an ISAR measurement can be written as
\begin{equation}
\label{eq:inv_prob_eq}
\vec{A} \vec{x} = \vec{y},
\end{equation}				
where $ \vec{A} $ is a  forward operator that propagates the isotropic, in frequency and angle, point sources on the defined grid to the observed RCS as a function of frequency and angle. $ \vec{x} $ is the image and $ \vec{y} $ is the measured RCS amplitude. $ \vec{A} $, $ \vec{x} $, and $ \vec{y} $ are all complex-valued. The RCS, $\sigma$, and the RCS amplitude, $\vec{y}$, are related by,
\begin{equation}
	\label{eq:RCS_2} 
	\sigma [\mrm{m^2}] = |\vec{y}|^{2}.
\end{equation} 

The image data, $ \vec{x} $, in (\ref{eq:inv_prob_eq}) can be determined directly using the backprojection method which is a standard method for ISAR imaging\cite{Vaupel+Eibert2006,Larsson2014,Larsson2016b}. The adjoint of $ \vec{A} $, $\vec{A}^{\dagger}$, is used to find a solution for $ \vec{x} $ using 
\begin{equation}
\label{eq:bp_eq}
\vec{A}^{\dagger} \vec{y} = \vec{x}^{\mrm{BP}},
\end{equation} 
where $\vec{x}^{\mrm{BP}}$ is the backprojection estimate of $\vec{x}$. \rev{The main advantage of backprojection is that it is straightforward to implement and robust.}

The target scattering contributions in these measurements are in most cases inherently sparse and (\ref{eq:inv_prob_eq}) can therefore be solved efficiently using an $ \ell_1 $-minimization method. Basis Pursuit Denoise (BPDN) as implemented in SPGL1 \cite{spgl1site,Berg+Friedlander2008} is used to solve (\ref{eq:inv_prob_eq}) in this paper. The point scattering model is defined in the forward operator $ \vec{A} $ and the backward operator, $ \vec{A} $'s conjugated transpose, $\vec{A}^{\dagger}$. The problem to be solved is formulated as
\begin{equation}
\label{eq:l1_minimization_eq}
\text{Minimize} \enspace || \vec{x} || _{1}  \enspace  \text{subject to} \enspace 
||  \vec{A} \vec{x} - \vec{y}|| _{2} \leq \kappa ,
\end{equation}
where the indices 1 and 2 denote the $ \ell_1 $ and $ \ell_{2} $-norms, respectively, and $ \kappa $ is an estimate of the misfit to the model. It is chosen to $ \kappa = 0.01 \cdot || \vec{y}|| _{2}$ for the simulations in this paper. Using BPDN described by (\ref{eq:l1_minimization_eq}) promotes sparsity in the solution for $ \vec{x} $\cite{spgl1site,Berg+Friedlander2008}. 

A sequence of reweighted $\ell_1$-minimizations operations that each are based on the preceding result are used. The purpose is to suppress the largest non-zero coefficients and increase the sensitivity to identify the smaller coefficients\cite{Candes+etal2008}.
The reweighting is refined by using smooth reweighted $\ell_1$-minimization \cite{Pinchera+etal2019,Pinchera+Migliore2021}. The procedure is adapted to ISAR image processing in \cite{Larsson+Gallstrom2022}. A smoothing function $\vec{D}$ is defined as
\begin{equation}
\label{eq:D}
  \vec{D} (\Delta r) = \begin{cases}
    1-(\Delta r/ d), & \text{if $\Delta r < d$}\\
    0, & \text{otherwise},
  \end{cases}
\end{equation}
where $\Delta r$ is the distance from the grid point. 
$\vec{D}$ is used to create $\vec{x}^D$ which is a smoothed version of $\vec{x}$
\begin{equation}
\label{eq:zD}
\vec{x}^D = |\vec{x}| \ast \vec{D},
\end{equation}
where $\ast$ denotes a 2D convolution between $|\vec{x}|$ and $\vec{D}$. Only grid points within a radius $d$ are used in the convolution. $\vec{x}^D$ has the same size as $\vec{x}$. $\vec{x}^D$ is then used to calculate a reweighting matrix $\vec{W}$ with its elements defined by
\begin{equation}
\label{eq:W}
\vec{W_{\mrm{i,j}}} = \frac{1}{\vec{x^D_{\mrm{i,j}}} + \eta},
\end{equation}
where $\eta$ is added to prevent division by zero. The requirement on $\eta$ is that it should be small and it is set to the smallest nonzero value of $\vec{x^D_{\mrm{i,j}}}$ \cite{Pinchera+Migliore2021}. The exact value does not affect the algorithm performance. The problem is reformulated using the reweighting matrix $\vec{W}$ as
\begin{equation}
\label{eq:l1_minimization_W_eq}
\mrm{Minimize} \enspace || \vec{W} \circ \vec{x} || _{1}  \enspace  \mrm{subject}  \enspace  \mrm{to}  \enspace 
||  \vec{A} \vec{x} - \vec{y}|| _{2} \leq \kappa ,
\end{equation}
where $\circ$ denotes the element by element multiplication of two matrices of the same size. This iterative procedure collects the points associated with a scattering center into a cluster. It converges fast and four iterations are used for all examples shown here. Good results are obtained for the smoothing function (\ref{eq:D}) with $d$ chosen to equal two times the grid spacing. 
The procedure can  be summarized in the following steps: 
\begin{enumerate}
\item{Determine $\vec{x}$ for the full data set by using smooth reweighted $\ell_1$-minimization (\ref{eq:l1_minimization_W_eq}). This is the first iteration}
\item{\label{itm:second}Use $\vec{z}$ to determine $\vec{z}^D$, with (\ref{eq:zD}), and  $\vec{W}$ }, with (\ref{eq:W})
\item{Determine a new $\vec{z}$ by using (\ref{eq:l1_minimization_W_eq})} 
\item{Go to step \ref{itm:second} or stop if the maximum number of steps is reached}
\end{enumerate}

\section{Results and Discussion}
\begin{figure}
\centering
\includegraphics[width=0.9\columnwidth]{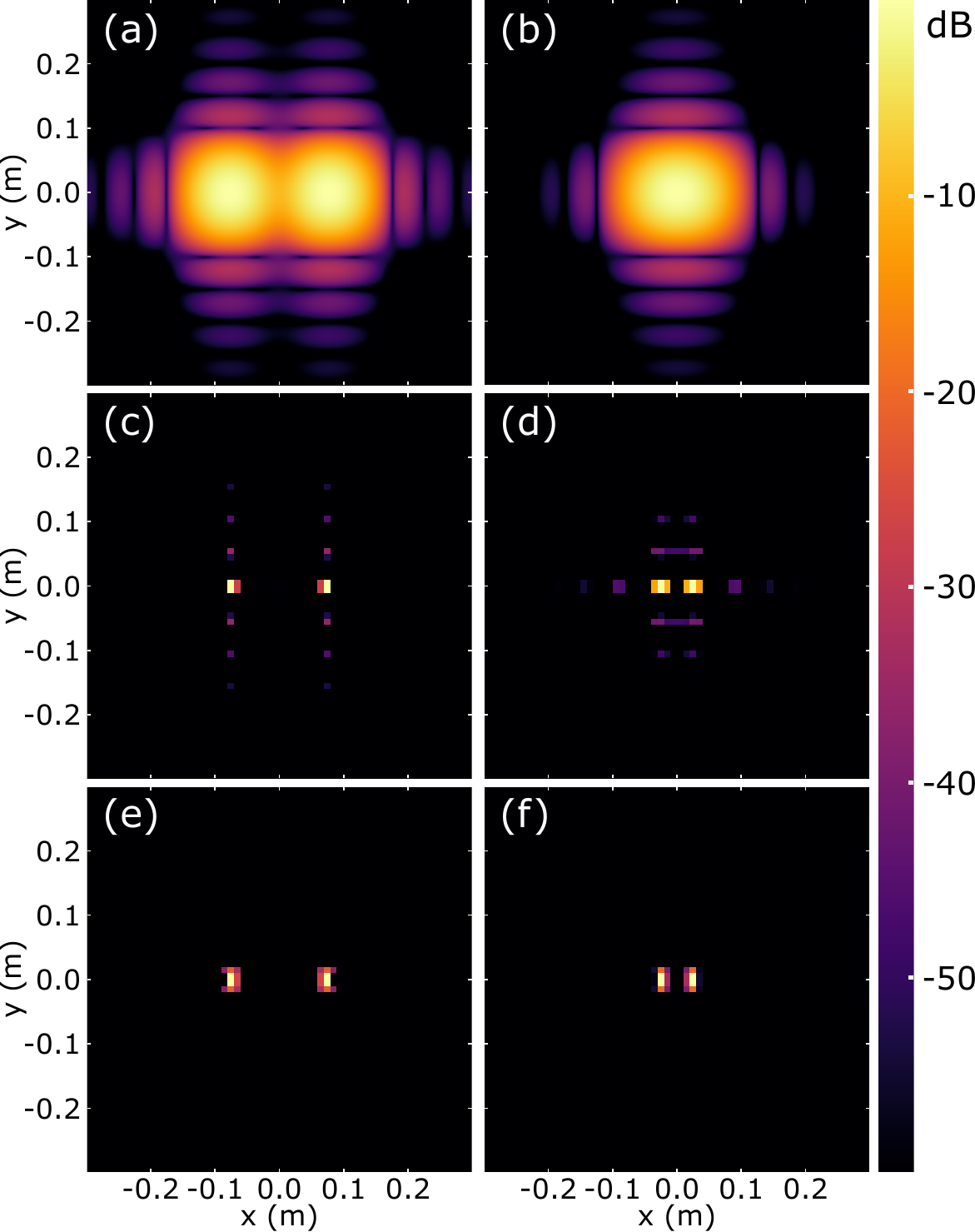}
\caption{Two point sources with equal amplitude, separated by 15\,cm in the left column and 5\,cm in the right column  are imaged using different methods. All images are normalized to a maximum of 0\,dB. (a)-(b) show backprojection with a Hann window, (c)-(d) show $ \ell_1 $-minimization results for the point sources, and (e)-(f) show iterative smooth reweighted $ \ell_1 $-minimization results for the point sources.}   
\label{fig:twopoints}
\end{figure}

\begin{figure}
\centering
\includegraphics[width=0.9\columnwidth]{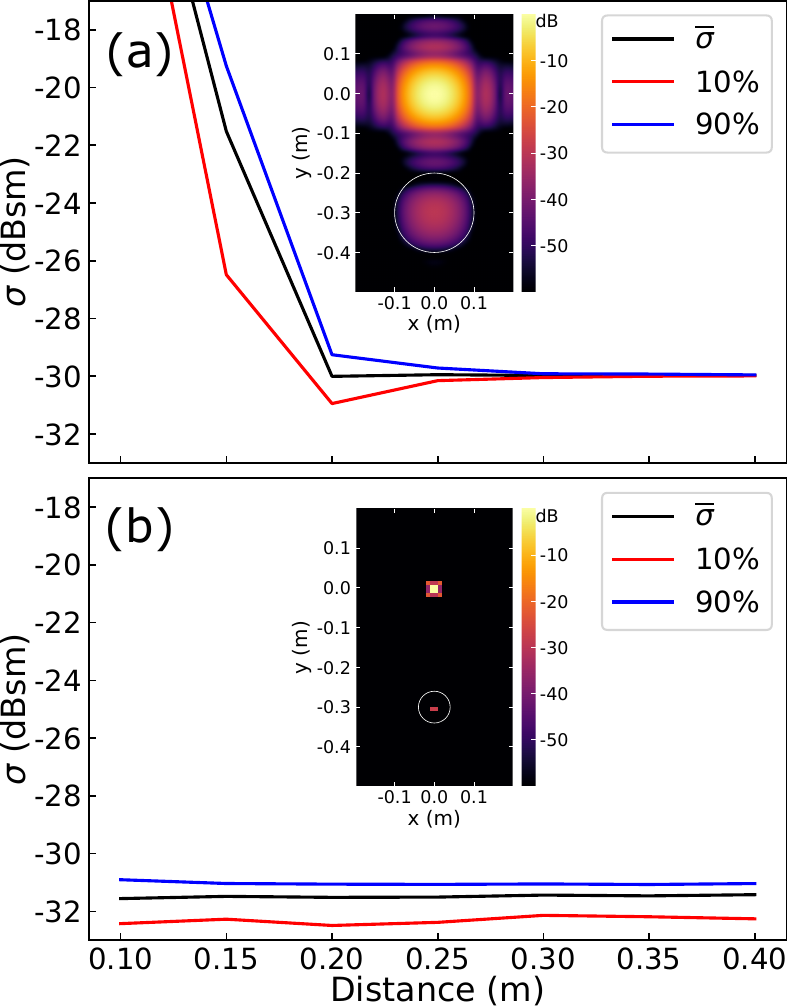}
\caption{Extraction of the RCS for a $-$30\,dBsm point source as a function of the distance from a 0\,dBsm point source at the origin for back projection. Frame (a) shows the average, the 10\ts{th} percentile, and the 90\ts{th} percentile for the extracted RCS from imaging using backprojection. One of the backprojection images is inserted. Frame (b) shows the same quantities using ISR.}   
\label{fig:stat}
\end{figure}

Farfield simulated results for two isotropic point scatterers with the same amplitude are given in Fig.~\ref{fig:twopoints}. Two cases where the point scatters are separated by 0.15\,m and 0.05\,m, respectively are shown for three different imaging methods. A 13.5--16.5\,GHz frequency range in 41 frequency steps and -5.7\degree--\, 5.7\degree azimuth angle range in 41 angle steps are used as examples to illustrate the methods in this section. The azimuth angle range is chosen so that the cross range resolution matches the down range resolution for backprojection. The down range resolution is 5\,cm as defined by $\clight / (2B)$, where $\clight$ is the speed of light and $B$ is the frequency bandwidth. The grid spacing is 1\,cm for all simulations.  Frame (a) and (b) show the results using backprojection with a Hann window. The resolution is limited by the used frequency bandwidth and azimuth angle range. Frame (c) and (d) show the results using $ \ell_1 $-minimization. This gives a much improved resolution compared with backprojection. However, the grid mismatch causes the points to spread out giving something similar to a sidelobe structure. The result are improved using ISR $ \ell_1 $-minimization as shown in frame (e) and (f) with the two point scatters represented by point clusters. This is an example of so called cluster sparsity as defined in \cite{Pinchera+Migliore2021}.
An isotropic point source with an RCS of {$-$30\,dBsm}, or 0.001\,m$^2$, at a variable distance from a point source at the origin with an RCS of 0\,dBsm, or 1\,m$^2$, is simulated with the same parameters as in the previous example. The aim is to extract the RCS of the the point source. This is done at 360 different positions in 1$\degree$ increments to obtain statistics. Fig.~\ref{fig:stat} shows an example where the  {$-$30\,dBsm} source is 0.3\,m from the origin. 
The result of the image gating RCS extractions for backprojection is shown in frame (a) and for ISR $ \ell_1 $-minimization is shown in frame (b) in Fig.~\ref{fig:stat} . An example of ISAR images are shown as inserts for the two methods. The RCS value was evaluated for 15\,GHz. The graphs show the average, the 10\ts{th} percentile, and the 90\ts{th} percentile of the extracted RCS for the $-$30\,dBsm point source. It is clear that image gating using the backprojection images gives the most accurate and robust estimate of the RCS for distances between the two scatterers that are more than 0.2\,m for this case. The RCS extraction using ISR $ \ell_1 $-minimization works for shorter distances but the average value is offset from the correct value and the uncertainty as estimated from the percentile curves is large.  
\section{Conclusion}
We have found that iterative smooth reweighted $ \ell_1 $-minimization gives images with better resolution than backprojection images.  The improved resolution makes it possible to do RCS extraction for scattering sources with less separation than in backprojection. However, the accuracy of the extracted RCS for ISR $ \ell_1 $-minimization can be lower than the RCS extracted using backprojection. The conclusion is that backprojection is more robust and should be the preferred method except in those cases when the scattering features cannot be resolved in the backprojection image where ISR $ \ell_1 $-minimization should be used.

The presentation will give examples of RCS extraction from measured RCS data and discuss the pros and cons of the different methods for ISAR.

\bibliographystyle{IEEEtran}
\bibliography{references}
\end{document}